# Improving Resolution and Run Outs of a Superconducting Noncontact Device for Precision Positioning

I. Valiente-Blanco, E. Diez-Jimenez, J. Sanchez-Garcia-Casarrubios, and J. L. Perez-Diaz

The authors are with the Departamento de Ingeniería Mecánica, UniversidadCarlos III de Madrid, Leganes ES-28911, Spain (e-mail: ivalient@ing.uc3m.es; ediez@ing.uc3m.es; jusanche@ing.uc3m.es; jlperez@ing.uc3m.es).

***Abstract*—This paper describes the improvement of the performance of a long-stroke high-precision positioning slider suitable for cryogenic environments due to the application of a set of design rules given in a previous paper. The device, based on superconducting magnetic levitation, is self-stable and does not make contact between the slider and the guideline, avoiding tribological and lubrication problems associated with cryogenics. This new prototype was built and tested in a relevant environment at 15 K and high vacuum ($\sim 10^{-6}$ Pa), demonstrating an enhanced resolution ($70 \pm 10$ nm RMS), lateral run out (about $\pm 2$ μm), and angular run outs (between tens to hundreds of μrad). In addition, new data related to the dynamics of the mechanism are presented. The demonstration of the design rules for this sort of mechanism provides a probed useful tool for engineers and increases the readiness of the technology. The demonstrated performance of the mechanism makes it suitable for applications where high-precision positioning is required over a long range in cryogenic environments like in far-infrared interferometry.**

## I. Introduction

Ultraprecision positioning has become an important development for precision mechanisms and devices. Precision positioning in a long stroke is an increasingly demanded requirement for applications both at room temperature [1]–[4] and in cryogenic environments as is the case of refocusing mechanisms, optical path difference actuators (OPDA), cryogenic motors, multiaxis positioning stages, conveyors, and actuators [5], [6]. Infrared interferometer spectroscopy [7] is an application of particular interest for the device [8]. Some typical requirements for these applications are: resolution $<1$ μm, step $<750$ nm, temperature 20 K, lateral deviations $\pm 100$ μm, $\pm 45$ μrad, and peak power consumption less than 10 mW. Two main solutions for high-precision positioning are found in the literature: piezoelectric actuators and active magnetic levitation mechanisms.

Piezoelectric actuators [9]–[11] have several limitations. For example, their range of motion is usually limited to a few hundred micrometers [12]; are sensitive to environmental changes such as temperature [13]; require high voltage to operate [14]; and exhibit hysteresis, nonlinearities, and drift [15], [16]. Dual-stage systems can cover longer strokes with impressive accuracy, also in cryogenic environments [17], but present problems like track-seek or track-follow [2].

Active magnetic devices are contactless systems that eliminate tribological problems related to contact reducing durability and fatigue problems at the same time [2], [6]. In comparison with the piezoelectric positioners, active magnetic levitation systems can offer similar accuracy for a larger motion range [18], [19]. However, they require an active system to establish and control levitation and position (frequently coils) as they are naturally unstable as stated by Earnshaw's theorem [20]. This leads to a larger electrical consumption, increased control complexity, and reduced reliability. These facts are undesirable in cryogenic and space applications.

Superconducting levitation has been previously proposed to solve these problems of active magnetic systems [21]. However, strokes not longer than a few mm [22], and typical resolutions of a few μm have been reported [23], [24]. In previous works [25], [26] the authors have presented a novel concept of a superconducting high-precision positioner with unprecedented performance. Due to the inherent stability of passive superconducting levitation [27]–[30] the slider levitates stably. Open-loop control of the current in the coils is required just to control the position of the slider in the sliding direction, but not to establish or stabilize levitation. This makes it possible to reduce the number of sensors, reduce the complexity of the control and power consumption, improves its reliability, and reduces supply and cooling requirements with respect to active maglev systems [2], [4]. In addition, the device is resistant against electrical shutdowns.

In a recent publication [31], a set of design rules for this sort of mechanism has been presented. Those design rules have been implemented to design and build an improved prototype with enhanced performance. The improved prototype was tested under cryogenic conditions, providing a demonstration of the design rules. In addition, new data about the dynamic behavior of the device are presented.

## II. Device Description

The device presented in this paper is an improved prototype of a previous device [26]. It is composed of three main parts: guideline, slider, and actuating system.

The stator is made of two 45-mm diameter superconducting polycrystalline $YBa_2Cu_3O_{7-x}$ disks (HTS), and an additional third 36-mm diameter central disk (Fig. 1). The slider is composed of a 160-mm length $Nd_2Fe_{14}B$ permanent magnet (PM) (2), magnetized vertically, carrying an optical mirror cube (6), and levitating over the superconductors [see Fig. 1(b)].

Due to the translational symmetry of the magnetic field generated by the PM [25], a sliding kinematic pair is established between the PM and the HTS. Thus, the slider is able to move in the sliding degree of freedom (DoF) with very low resistance. Greater restoring forces appear if the PM is moved in any other direction due to breakage of the symmetry of the magnetic fields applied to the HTS.

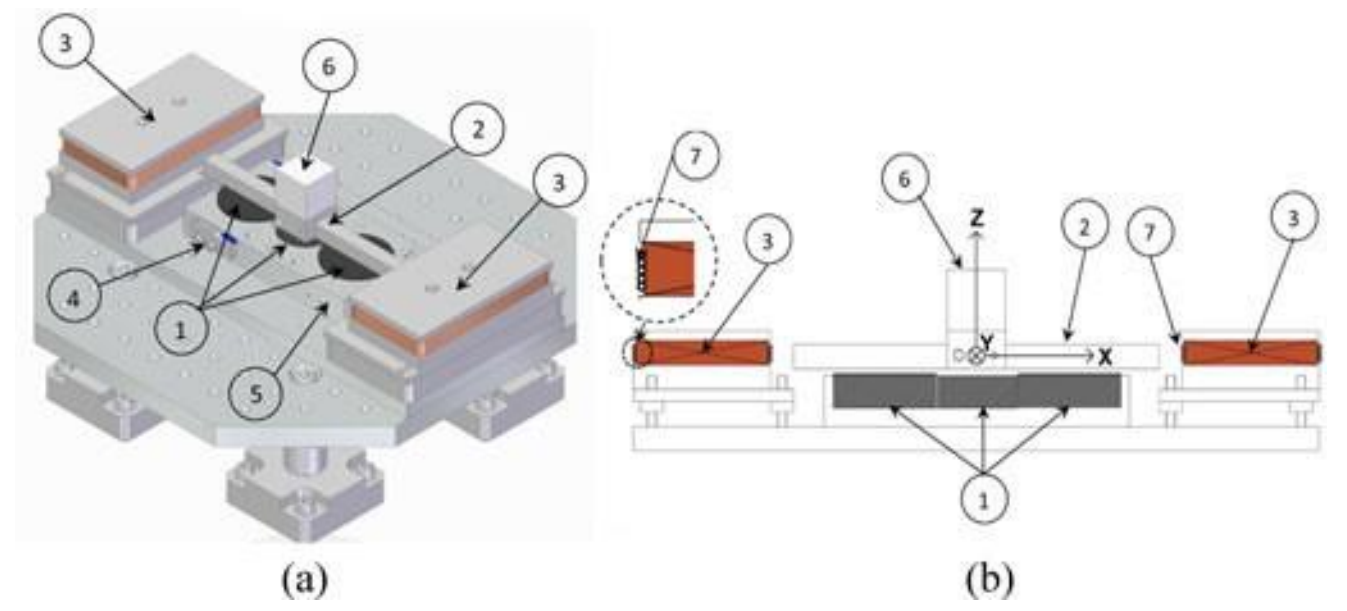

Fig. 1. (a) CAD representation of the device; (b) lateral representation + reference system: (1) YBaCuO superconductor disks (HTS), (2) long permanent magnet (PM), (3) aluminum core coils, (4) PT-100 sensors, (5) HTS aluminum vessel, (6) aluminum polished optic cube, and (7) fine positioning coils.

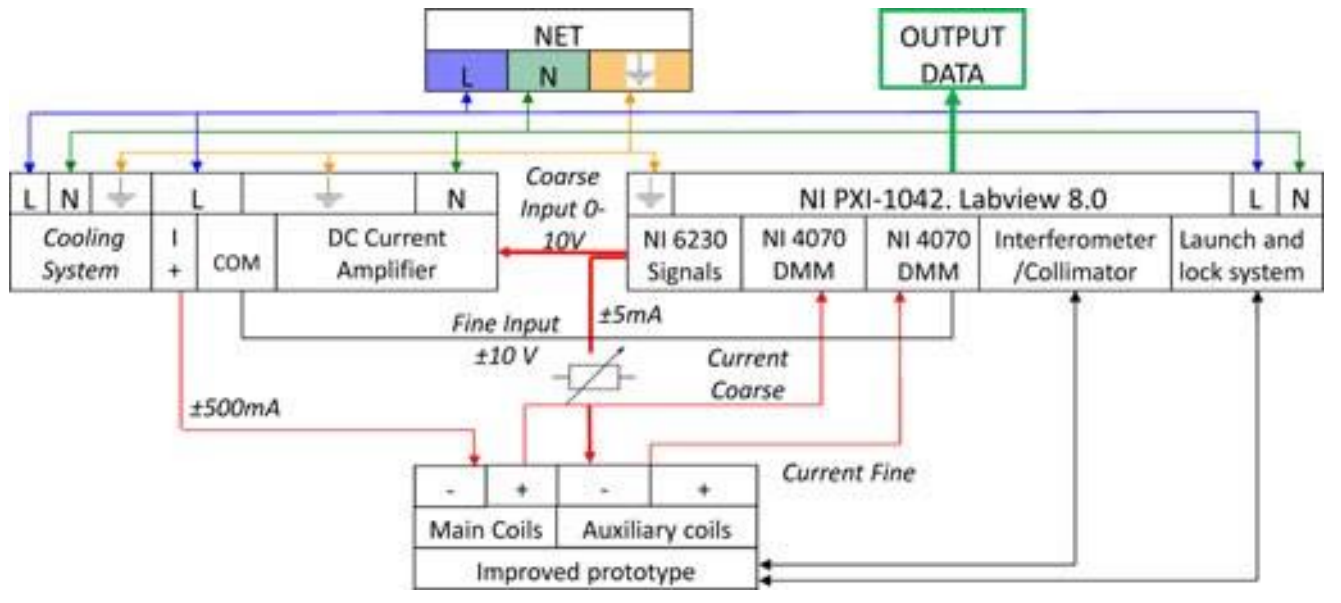

Fig. 2. Electrical scheme of the control system.

Finally, a set of coils placed at both ends of the stroke of the slider apply contactless magnetic forces to the slider in order to control its position at distance and without contact.

### A. Stator Modifications

Including a third disk at the center of the guideline increases stiffness to the motion in the constraint directions (reducing run outs) without increasing the resistance to the slide. This is true, provided magnetic field variations in this third disk are negligible for motion in the sliding direction [26].

The levitation height of field cooling (HFC) is the distance between the HTS upper surface and the bottom surface of the PM in the slider. In order to increase stiffness and reduce run outs, the HFC was reduced from 3 to 2 mm [32]–[34]. Because the resistance to the slide is expected to increase [31], the actuation system was improved to minimize the increase in the power consumption.

### B. Actuating System Redesign

Two sets of coils placed at either end of the stroke exert contactless magnetic forces in the slider in order to move it along the motion direction. The coils have a rectangular shape that minimizes run outs due to misalignments between the coils and the slider. The actuating system is divided in two subsystems. A first system is a coarse-step motion stage composed of the main coils (3) with several turns in a large area. Coils were redesigned with a larger number of turns than those in [26] to achieve a tradeoff between an improvement of the performance and the expected increase of the power consumption of the device mentioned in the previous section. A second system of fine-step motion is composed of two pairs of auxiliary coils with just a few turns (7) surrounding the main ones (3) exerting a lower force on the PM and, therefore, improving the resolution.

### C. Electrical Control System Configuration

Because of this new coarse plus fine actuating system, the electrical control system was also modified, leading to the final configuration shown in Fig. 2.

The position of the slider is controlled by two independent current signals (fine plus coarse). The fine current signal is directly provided by a NI 6230 16-bit card connected to a precision resistor. The current signal required for the coarse-step actuation system is provided by a voltage-controlled dc current amplifier which was specially designed to provide a greater stability of the output current and isolate power supply noise. In the end, the amplifier can provide ±500 mA current with a current accuracy of ±30 $\mu$A.

## III. Experimental Setup

The device was tested in a high vacuum ($2.7 \times 10^{-6}$ Pa) and at a temperature of around 15 K. Wire-wound PT-100 sensors [see (4) in Fig. 1(a)] were used to measure the prototype temperature. Both the PT-100 and the superconductor disks (1) were fixed to an aluminum base (5), as shown in Fig. 1(a). The distance between the revolution axes of the 45-mm diameter HTS disks was fixed at 84 mm. The small HTS was fixed with its axis coincident with the Z-axis in Fig. 1(b). The levitation height (HFC) was set to 2 mm.

Once the required environmental conditions were reached the PM was released levitating over the superconducting disks. Two independent current signals circulating through the coils modify the position of the slider. A NI 4070 FlexDMM multimeter, with 1 $\mu$A resolution and 10 ppm accuracy at full scale (1 A) was used to measure current and voltage signals.

In order to measure the position of the slider, a plane mirror interferometer, Agilent model 10706B, was used. A Newport collimator, model LDS-Vector (accurate to within 3%), measured the angular run outs of the slider. A polished aluminum optical cube (6) was attached to the slider, as shown in Fig. 1(a), to reflect the laser beam. A picture of the prototype inside the cryostat before cooling is shown in Fig. 3.

## IV. Results and Discussion

### A. Stroke, Stability, and Resolution

The position of the slider along the stroke of the mechanism (±7.5 mm) was measured for different current values.

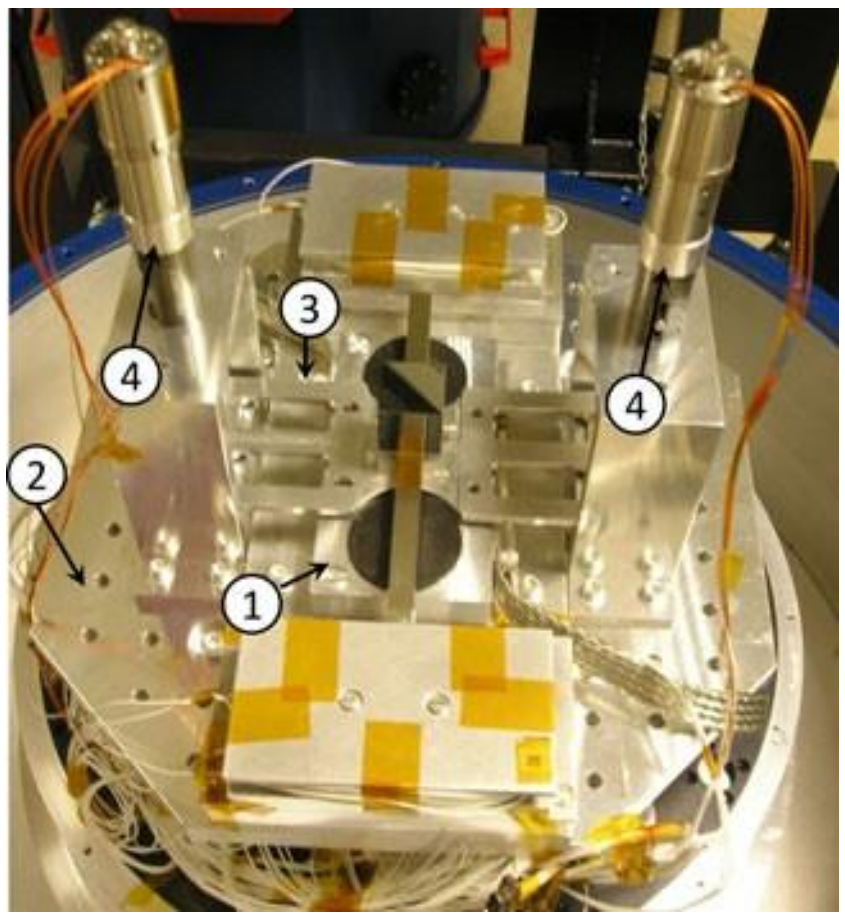

Fig. 3. View of the improved prototype (1), attached to the cryostat cold plate (2). The slider is held by the launch lock support (3), which will release the magnet using a couple of cryogenic linear actuators from LIDAX (4).

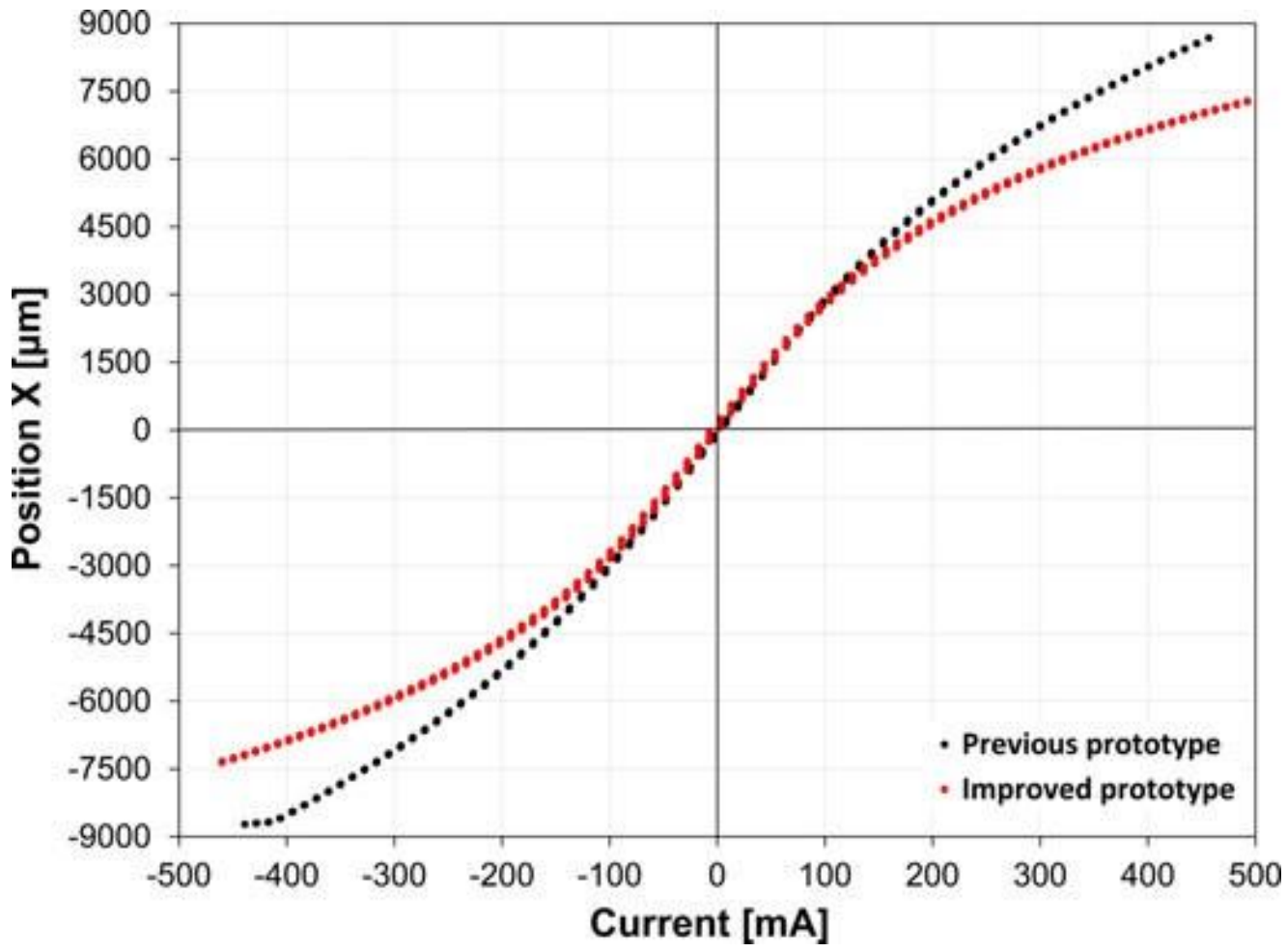

Fig. 4. X position of the slider versus current in the coils. X standard deviation equal to 0.4 $\mu$m, and current standard deviation equal to $\pm$50 $\mu$A.

Hysteresis for full stroke motion reached its maximum at the central position (X = 0) and is about 100 $\mu$m. The increased hysteresis can be explained by the increased magnetization changes in the HTS [26] due to the reduction of the HFC.

Taking into account the residual resistance ratio (RRR = 500) of the coils in the actuation system at 15 K, the peak power consumption was measured to be about 6 mW.

X position versus time around the central position (X = 0 mm) for an acquisition frequency of 1 kHz is shown in Fig. 5.

Position stability was measured at different points on the stroke of the slider. It can be assumed to be almost independent of the current signal and the slider position. However, it is influenced by external vibration sources such as the cryo-pump or structural vibrations in the lab building. Ultimately, the X position accuracy is about $\pm$0.4 $\mu$m for any X position along the path of the device. Y position stability was measured to be $\pm$0.4 $\mu$m too, almost independent of the total current in the coils.

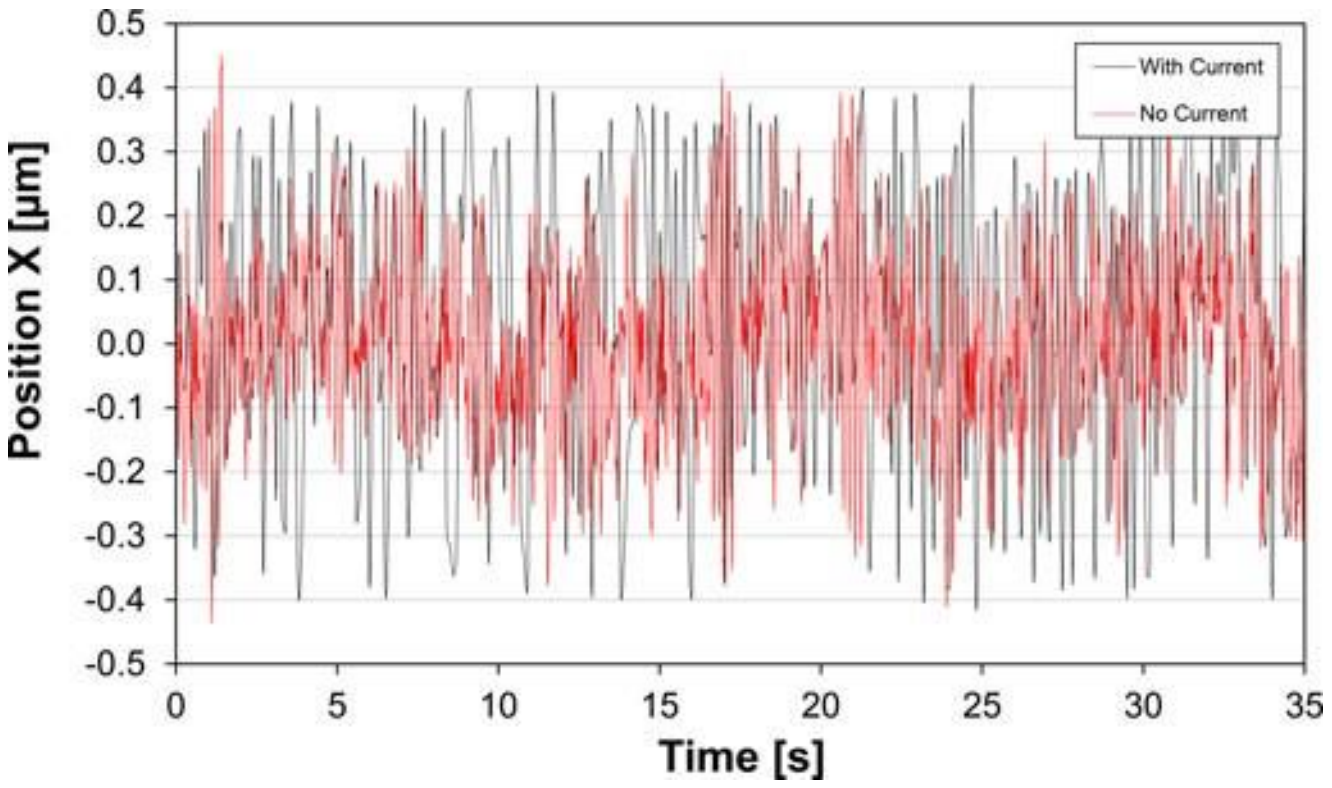

Fig. 5. Stability of the X position of the slider versus time in the central position (with and without current in the coils).

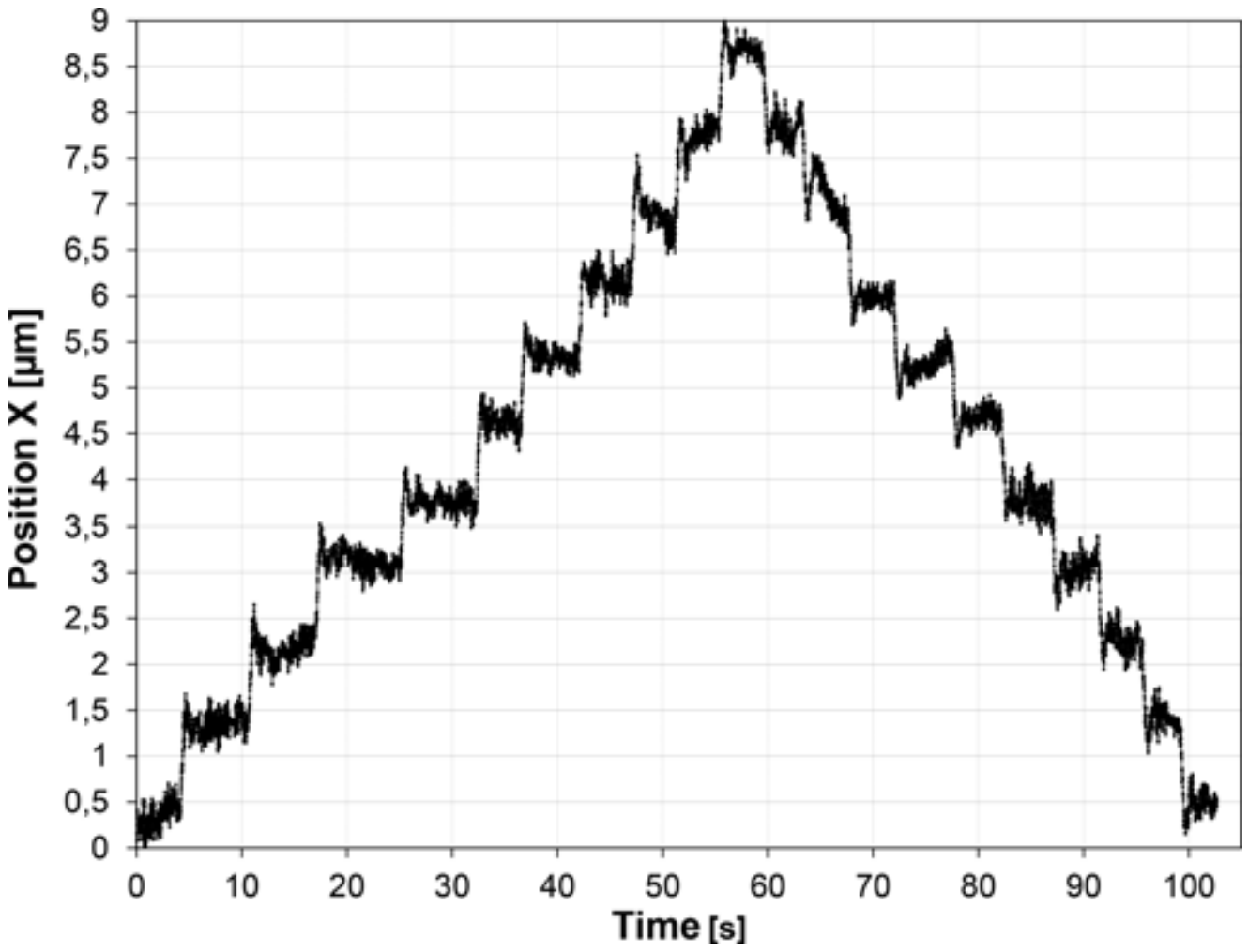

Fig. 6. X position of the slider versus time for current increment $\Delta C \approx 30\,\mu$A.

X position of the slider versus time for a set of steps in the current of 30 $\mu$A is shown in Fig. 6. The position resolution under this operating condition was calculated to be 0.8 $\pm$ 0.1 $\mu$m.

Fig. 7 shows the X position of the slider versus current in the auxiliary coils for a current increment of $\Delta C \approx 55 \pm 10\,\mu A$ near the central position (X = 0). X position resolution of 70 $\pm$ 10 nm was estimated from data shown in Fig. 8.

Both, the coarse and fine positioning resolutions were improved with regard to the previous prototype [26] (see Table I).

### B. *Lateral Run Out, Pitch, Yaw, and Roll*

Lateral run out (deviation in the Y-axis) versus X position of the slider is shown in Fig. 8. Pitch (rotation around the Y-axis in Fig. 1) and roll (X-axis) versus X position are shown in Fig. 9. Both were significantly improved due to the implementation of the design rules [31] to the redesign of the stator (increased resistance to the motion in the undesired directions), the actuation system (enhanced stability) and a better alignment of the PM in the slider and the coils provided by an improved launch

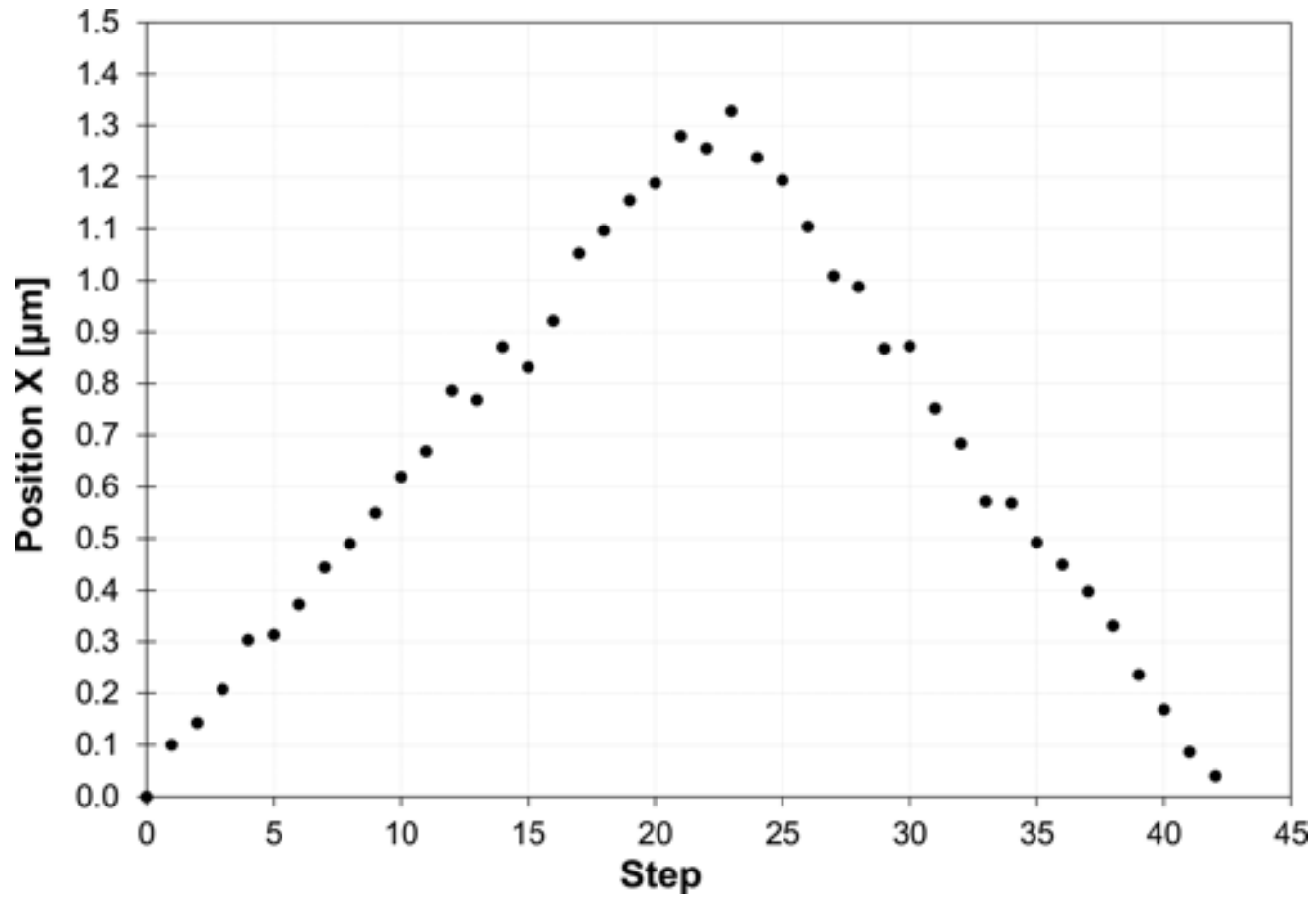

Fig. 7. X position (RMS) of the slider versus current step, for $\Delta C = \pm 55\ \mu$A. Samples measured consecutively, with no time register.

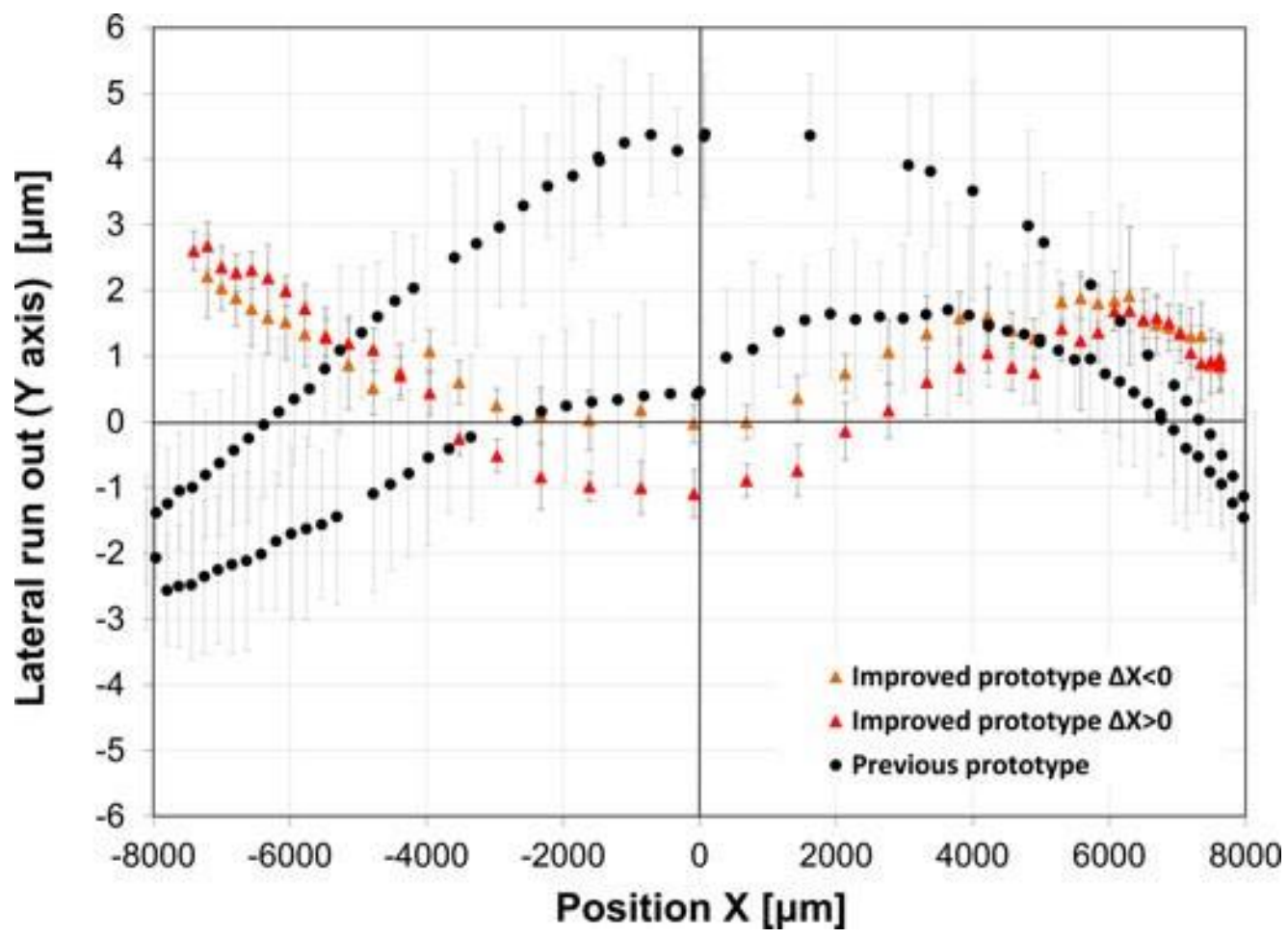

Fig. 8. Lateral run out versus X position of the slider. X accuracy = 50 $\mu$m.

and lock mechanism (see Fig. 2) that provides greater stiffness and enhanced alignment capability.

### *C. Dynamic Behavior*

Dynamic behavior for unforced oscillations was characterized for this device for the first time. The slider was moved away from the initial equilibrium position (X = 0 mm) to different positions in the stroke. Then, the current in the coils was switched off to the open circuit ($t$ = 1 s in Fig. 10). Results for this experiment are shown in Fig. 10.

The dynamic behavior fits well with an underdamped harmonic oscillator. Spectral analysis using the Lomb-normalized periodigram method reveals an average damped frequency of 0.93 $\pm$0.02 Hz that does not strongly depend on the motion amplitude within the stroke of the device. The damping ratio was calculated to be 0.37$\pm$0.01. Damping of the slider is associated to energy dissipation in the superconductors [35] and eddy current effects. Speeds up to 30 mm/s were measured in the dynamic test.

TABLE I
SUMMARY OF THE IMPROVEMENTS OF THE DEVICE

| | Previous prototype [26] | Improved prototype |
|---|---|---|
| Resolution: coarse [$\mu$m] | 1.5 ± 0.2 | 0.8 ± 0.2 |
| Resolution: fine [nm] | 230 ± 30 | 70 ± 10 |
| Positional stability of X [$\mu$m] | ±1 | ±0.4 |
| Lateral run out (Y-axis) [$\mu$m] | ±4 | ±2 |
| Relative pitch (Y-axis) [$\mu$rad] | ±650 | ±65 |
| Relative yaw (Z-axis) [$\mu$rad] | ±50 | ±60 |
| Relative roll (X-axis) [$\mu$rad] | ±2000 | ±350 |
| Stroke [mm] | ±9 | ±7.5 |
| Sensitivity of coarse motion [nm/$\mu$A] | 30 ± 1 [8 mm] | 29 ± 1 [6 mm] |
| Power consumption [mW] | 5 | 6 |
| Maximum hysteresis [$\mu$m] | 60 | 100 |
| Damped frequency ($\omega_d$) [Hz] | Not measured | 0.93 ± 0.02 |
| Damping ratio ($\xi$) [Hz] | Not measured | 0.37 ± 0.01 |
| Maximum speed [mm/s] | Not measured | 30 ± 3 |

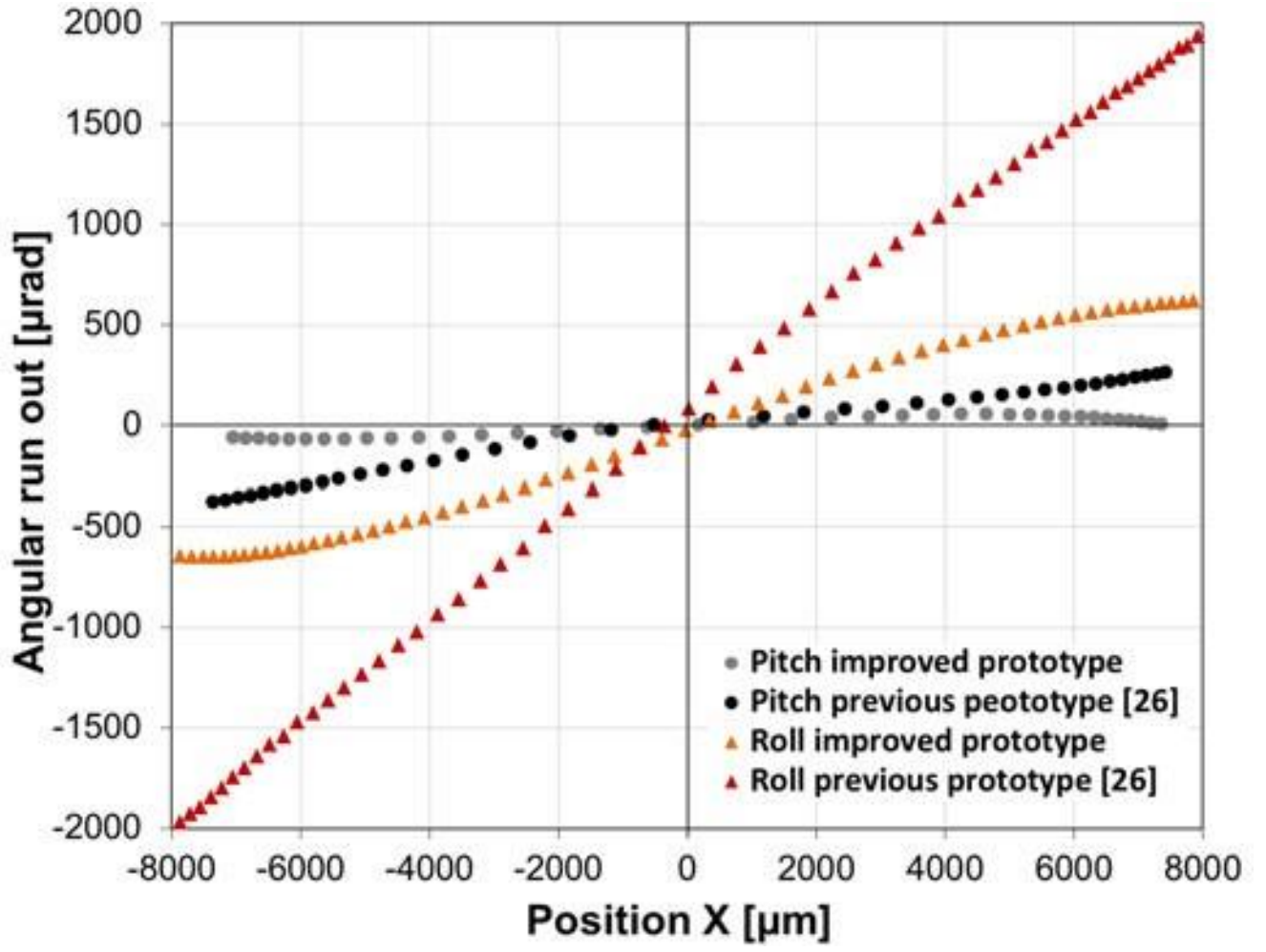

Fig. 9. Angular run outs (pitch and roll) versus X position. X accuracy = 50 $\mu$m.

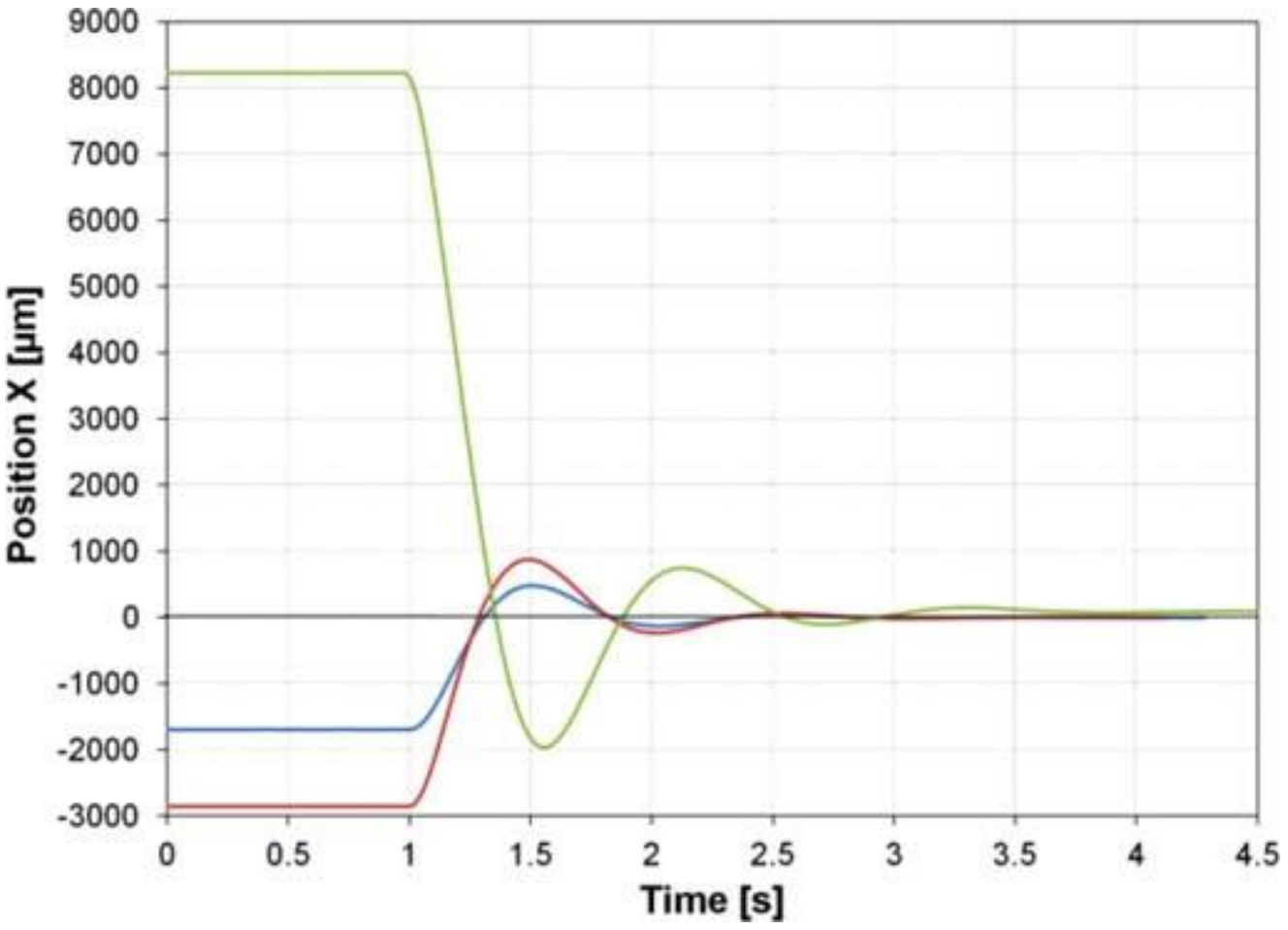

Fig. 10. X position versus time for different initial positions. Coils were switched off at $t$ = 1 s in all cases.

## V. Conclusion

An improved prototype of a long-stroke precision positioner based on superconducting magnetic levitation has been presented. This prototype was designed and built looking for an improvement in the positioning resolution and run outs following the design rules provided in a previous paper [31]. Improvements were demonstrated under cryogenic conditions (15 K and $\sim 10^{-6}$ Pa): a resolution three times lower, a lateral run out two times lower, and angular run outs of about one order of magnitude lower than previous prototypes with a slightly increased power consumption of 6 mW. Finally, new measurements about the dynamic behavior of the device are presented. The improvements are detailed in Table I.

A demonstration of the utility of the design rules [31] is presented that improves the performance of this kind of superconducting devices [26] to a level that makes them attractive for applications such as far-infrared interferometry or cryogenic spectroscopy in which sensors must operate at very low temperatures.

## Acknowledgment

This work was supported in part by Dirección General de Economía, Estadística e Innovación Tecnología, Conserjería de Economía y Hacienda, Comunidad de Madrid, ref. 12/09

The authors offer their heartfelt thanks to the LIDAX team for their technical support and cooperation in this project.